\magnification=1200 \baselineskip=13pt \hsize=16.5 true cm \vsize=20 true cm
\def\parG{\vskip 10pt} \font\bbold=cmbx10 scaled\magstep2

\centerline {\bbold Broad Histogram Relation Is Exact}\parG
\centerline{Paulo Murilo Castro de Oliveira}\par
\centerline{Instituto de F\'\i sica, Universidade Federal Fluminense}\par
\centerline{av. Litor\^anea s/n, Boa Viagem, Niter\'oi RJ, Brazil
24210-340}\par
\centerline{e-mail PMCO @ IF.UFF.BR}\parG
to be published in {\it Eur. Phys. J.} {\bf B}

\vskip 1cm\leftskip=1cm \rightskip=1cm
{\bf Abstract}\par
The Broad Histogram is a method designed to calculate the energy degeneracy
$g(E)$ from microcanonical averages of certain macroscopic quantities $N_{\rm
up}$ and $N_{\rm dn}$. These particular quantities are defined within the
method, and their averages must be measured at constant energy values, i.e.
within the microcanonical ensemble. Monte Carlo simulational methods are used
in order to perform these measurements. Here, the mathematical relation
allowing one to determine $g(E)$ from these averages is shown to be exact for
any statistical model, i.e. any energy spectrum, under completely general
conditions.\par

We also comment about some troubles concerning the measurement of the quoted
microcanonical averages, when one uses a particular approach, namely the
energy random walk dynamics. These troubles appear when movements
corresponding to different energy jumps are performed using the same
probability, and also when the correlations between successive averaging
states are not adequately treated: they have nothing to do with the method
itself.\parG

\par
\leftskip=0pt \rightskip=0pt
\vfill\eject

Consider a statistical system at thermal equilibrium under constant
temperature $T$ (we set the Boltzmann constant to unit). The canonical
average reads

$$<Q>_T\,\, = \, {\sum_S Q_S \exp(-E_S/T) \over \sum_S \exp(-E_S/T)}\,\,\,\, ,
\eqno(1)$$

\noindent where both sums run over all states $S$ available for the system,
each one corresponding to an energy value $E_S$ and another value $Q_S$ for
the particular quantity $Q$ one is interested in (magnetizatiom, for
instance). A very useful way to determine such an average is through computer
simulational methods, particularly the so-called importance sampling
introduced 45 years ago in a seminal paper [1]. The general idea is to
construct a Markovian chain of states available for the system. Along this
chain, each new state is obtained by performing some random modification (a
movement) at the current state in hands. Instead of all possible states, one
uses only the finite sub-set obtained by this Markovian process, in order to
calculate an approximation for the true average (1). The exponential terms
appearing in (1), i.e. the Boltzmann averaging weights, are automatically
taken into account during the construction of the Markovian chain:
increasing-energy random movements, i.e. $S \to S'$ where $\Delta E = E_{S'}
- E_S > 0$, are accepted only with probability $\exp(-\Delta E/T)$.

The value obtained from this recipe could be a very bad approximation, if the
user fails in taking into account some fundamental precautions. For instance,
consider an Ising model where $N$ spins can point up or down, corresponding
to $\Omega = 2^N$ possible states. In real implementations one has, say, $N
\sim 10^4$, and the number $M$ of states along the Markovian chain is always
much less than $\Omega > 10^{3,000}$. Thus, the tiny sub-set with $M$
averaging states may be a biased sample not representing the huge set with
$\Omega$ states. The simplest protocol one can adopt to generate a new state
from the current one is the single-spin-flip protocol: one chooses randomly
one spin, and flips it or not, according to the energy rules described above.
By repeating this single-spin-flip process $M$ times ($M \approx N = 10^4$),
one gets successive states very similar to each other (at most one different
spin among $N$), which would be statistically correlated. These correlations
will introduce systematic errors. Nobody follows this fool protocol. Indeed,
one normally considers a new state along the Markovian chain only after at
least $N$ spin-flips were tried, i.e. only after a whole-lattice sweep was
performed. Another precaution to be taken into account is to discard a
certain number of initial states along the Markovian chain, in order to avoid
possible biases caused by any oddness of the very first tossed state. This
corresponds to wait for a thermalization transient time.

In order to get the average $<Q>_T$ as a function of $T$, one must repeat the
computer run again and again, for many different fixed values of $T$. This is
a waste of computer time, compared to another possible method where the whole
temperature spectrum could be sampled in just one computer run. Indeed, still
following the same recipes, such a method was shown to be possible [2,3], at
least in principle. The idea is to perform an analytic reweighting of
equation (1), allowing one to obtain the average $<Q>_{T'}$ for a new
temperature $T'$, without performing a new computer run. In order to see how
this ``magic'' would be possible, we need first to introduce the
microcanonical average

$$<Q(E)>\,\, = \, {\sum_{S(E)} Q_S \over g(E)}\,\,\,\, ,
\eqno(2)$$

\noindent for the same quantity $Q$. Now, the sum runs over all states $S(E)$
belonging to the same energy level $E$. The energy degeneracy $g(E)$ counts
their number. Instead of keeping the system under a constant temperature $T$,
i.e. in contact with an equilibrated heat bath, now the condition is more
restrictive: the system is completely isolated from the environment, and its
energy is kept constant. Each state $S(E)$ within this energy level enters
equally weighted into this microcanonical average. Nevertheless, the same
kind of fundamental precautions described above must be followed by the user
who tries to approximate this average by random sampling some sub-set of the
whole $g(E)$ states.

The terms in both sums appearing in (1) can be arranged in groups
corresponding to the same energy, and the canonical average reads now

$$<Q>_T\,\, = \sum_E <Q(E)> P_T(E) \,\,\,\, ,
\eqno(3)$$

\noindent where

$$P_T(E) =  {g(E) \exp(-E/T) \over \sum_{E'} g(E') \exp(-E'/T)} \,\,\,\, .
\eqno(4)$$

\noindent The temperature $T$ appears only in the Boltzmann weights, the
degeneracies $g(E)$ being independent of $T$. One can easily express
$<Q>_{T'}$ in terms of $<Q>_T$ through

$$P_{T'}(E) = {{P_T(E) \exp[E({1 \over T}-{1 \over T'})]} \over
{\sum_{E'} P_T(E') \exp[E'({1 \over T}-{1 \over T'})]}} \,\,\,\, .
\eqno(5)$$

\noindent Thus, one needs a single computer run at some fixed temperature
$T$, measuring the probability distribution $P_T(E)$. This measurement can be
performed by accumulating the number of visits to each energy level $E$ on a
histogram. During the same computer run, one can also accumulate the
successive values of $Q_S$ in another $E$-histogram, allowing the
determination of $<Q(E)>$ which also is, of course, independent of $T$. Then,
the average $<Q>_{T'}$ can be obtained by using first equation (5) and then
replacing $T$ by $T'$ in (3). This is the essence of the so-called
reweighting methods [2,3].  Unfortunately, the probability distribution
$P_T(E)$ is very sharply peaked around the average energy $<E>_T$, and its
numerical evaluation is accurate only inside a tiny region around this peak.
Unless $T'$ and $T$ are very near to each other, the overlap between $P_T(E)$
and $P_{T'}(E)$ is negligible, and equation (5) is useless. The new peak
position $<E>_{T'}$ corresponds to the vanishing tail of the actually
measured distribution $P_T(E)$, where the statistics is poor. Thus, one has
no good accuracy at all near the peak of the inferred distribution
$P_{T'}(E)$. The larger the system size the worse becomes this problem, due
to the probability distribution sharpness.

That is why, in spite of the nice reasoning, reweighting methods have
difficulties in practice --- see, for instance [4]. Nevertheless, the
obvious but fundamental observation that both $g(E)$ and $<Q(E)>$ do not
depend on the particular temperature $T$ adopted in the computer
simulation remains an important information. Yet more fundamental is the
observation that both $g(E)$ and $<Q(E)>$ do not depend on any
thermodynamic constraints, that they are characteristics of the energy
spectrum alone and not of the particular interactions between the system and
its environment. Thus, in principle, it is possible to devise some computer
simulational method allowing the direct determination of these quantities.
Concerning the degeneracy $g(E)$, many such methods were tried --- see, for
instance [5] --- all of them relying on the histogram of visits to each
energy level.

The Broad Histogram Method [6] differs from all other methods I know. It
relies on the determination of $g(E)$ from the microcanonical averages, i.e.
equation (2), of two particular macroscopic quantities also introduced in
[6]. First, let's consider some protocol of allowed changes (movements) to be
performed at the current state. For instance, one can consider all possible
single-spin-flips for the case of an Ising system. Alternatively, one can
think about all two-spin-flips, to flip entire blocks containing up to $n$
spins, or any other protocol. Also for other models than Ising's, one can
previously determine to adopt any protocol of allowed movements. The only
important point to be noted is the microreversibility of such a protocol,
i.e. if some movement transforming state $S$ into state $S'$ is allowed by
the particular protocol, then the back transformation of $S'$ into $S$ is
also allowed, independent of probabilities. Consider two energy levels $E$
and $E+\Delta{E}$. Starting from a given state $S$ with energy $E$, the
number $N_{{\rm up},S}^{(\Delta{E})}$ counts all possible movements
increasing its energy by $\Delta{E}$. One needs to consider all possible $S'$
which can be achieved from the fixed current $S$, provided the movement ($S$
to $S'$) is allowed by the previously adopted protocol, and the energy jump
is $\Delta{E}$. Now, considering all the $g(E)$ states belonging to level
$E$, the total number of possible movements increasing the energy from $E$ to
$E+\Delta{E}$ is

$$\sum_{S(E)} N_{{\rm up},S}^{(\Delta{E})} = g(E) <N_{\rm up}(E)> \,\,\,\, ,
\eqno(6)$$

\noindent where the definition (2) of microcanonical average was used.
Analogously, starting from some state $S'$ with energy $E+\Delta{E}$,
$N_{{\rm dn},S'}^{(\Delta{E})}$ is the number of possible movements
decreasing its energy to $E$. The total number of possible movements from
level $E+\Delta{E}$ to level $E$ is

$$\sum_{S'(E+\Delta{E})} N_{{\rm dn},S'}^{(\Delta{E})} =
g(E+\Delta{E}) <N_{\rm dn}(E+\Delta{E})>
\,\,\,\, .
\eqno(7)$$

\noindent Due to the quoted microreversibility, these two numbers are equal,
and one has

$$g(E) <N_{\rm up}(E)> \, = \, g(E+\Delta{E}) <N_{\rm dn}(E+\Delta{E})>
\,\,\,\, ,
\eqno(8)$$

\noindent which is the fundamental relation introduced in reference [6]. It
is now proven to be exact for any statistical model, i.e. any energy spectrum
$g(E)$. Note that in both averages $<N_{\rm up}(E)>$ and $<N_{\rm dn}(E)>$
only movements corresponding to energy jumps $\Delta{E}$ and $-\Delta{E}$
respectively must be taken into account.

For any system, relation (8) can be used in order to determine $g(E)$ from
the microcanonical averages $<N_{\rm up}(E)>$ and $<N_{\rm dn}(E)>$
measured as functions of the energy $E$. These measurements (including
also $<Q(E)>$) can be performed by any Monte Carlo approach, the result's
accuracy depending exclusively on the quality of this particular approach
--- not on equation (8) which is exact. Once these two averages are known,
one can determine all the spectrum $g(E)$ from the ground state degeneracy
$g(E_0)$ supposed to be previously known. In practice, this previous
knowledge is not necessary (except for entropy estimates) because $g(E_0)$
cancels out in equation (4). Thus, the Broad Histogram Method consists in
measuring the microcanonical averages $<N_{\rm up}(E)>$, $<N_{\rm dn}(E)>$
and $<Q(E)>$ by using any Monte Carlo approach, the results being stored in
$E$-histograms. After that, when the computer simulation is already over,
equation (8) allows one to determine $g(E)$, and then equations (4) and (3)
can be used in order to determine the canonical averages of interest, for any
temperature $T$.

Equation (8) can be put into alternative forms. Taking $\Delta{E} << E$, one
can approximate it by

$${{\rm d} \ln g(E) \over {\rm d}E} = {1 \over \Delta{E}}\,\,
\ln {<N_{\rm up}(E)> \over <N_{\rm dn}(E)>}\,\,\,\, .
\eqno(9)$$

\noindent Moreover, by using a dirty mathematical transformation, one can
write also

$${{\rm d} \ln g(E) \over {\rm d}E} =
\ln {<\sum_{\Delta{E}} N_{\rm up}(E)^{1/\Delta{E}}>
\over <\sum_{\Delta{E}} N_{\rm dn}(E)^{1/\Delta{E}}>} \,\,\,\, ,
\eqno(10)$$

\noindent which could be useful in order to average various values of $\Delta
E$ simultaneously, saving computer time. Nevertheless, the dirty trick of
introducing the exponent $1/\Delta{E}$ inside the average brackets could
lead to systematic errors which remain to be verified by the user for each
case. For the Ising model in two and three dimensions, for instance, it works
very well [7].

Hereafter, we will discuss a particular Monte Carlo approach originally
adopted [6,7] in order to calculate the microcanonical averages used
within the Broad Histogram Method. Canonical simulations under a fixed
temperature $T$ cover only a narrow energy window around the average
value $<E>_T$. In order to obtain the microcanonical averages appearing
in (8), this is not a good strategy, because one needs to sample a broad
energy range. It does not help much to increase the computer time improving
the statistics on the exponentially vanishing tails of the distribution
$P_T(E)$: its width does not depend on the computer time. One possible
solution is to replace such canonical dynamics by a random walk along the
energy axis. This idea was implemented [6] by using a simple rejection rule:
any increasing-energy tossed movement is performed only with probability
$N_{{\rm dn},S}^{(\Delta{E})}/N_{{\rm up},S}^{(\Delta{E})}$, where $N_{{\rm
dn},S}^{(\Delta{E})}$ and $N_{{\rm up},S}^{(\Delta{E})}$ are measured at the
current state $S$, both corresponding to the same energy difference
$\Delta{E}$ of the tossed movement. Following this rule, the range of visited
energies will increase proportionally to $\sqrt{t}$, where $t$ is the
computer time. Thus, to obtain a broad energy histogram is a simple matter of
having enough computer time, within this RW dynamics (for random walk).
Actually, I discovered later that RW dynamics was previously introduced in a
nice paper [8], considering the much more general problem of optimization in
conflicting-interaction systems. Nevertheless, the dynamics introduced in [8]
is distinct from that introduced in [6] in a subtle but fundamental detail
discussed later.

The RW dynamics solves the problem of obtaining broad histograms. However,
there is no free lunch, and this advantage has a price: the correlations
appearing between successive states along the Markovian chain are worse to
treat than they are within canonical, fixed temperature dynamics, for which
correlations can be eliminated simply by waiting some few whole-lattice
sweeps before computing a new averaging state. In this case, this simple
precaution is enough, because the energy never jumps very far from the
average value $<E>_T$, thus the successive states are always thermalized (of
course, after the initial discarded transient steps). Within the RW dynamics,
on the contrary, big energy jumps occur, and just after one of them the
current state is not yet thermalized: it carries biases from the recently
visited far-away energies, and must be considered as an odd state concerning
this new energy level. The solution is not simply to wait for more spin
flips, because the energy would jump again and again. One possible solution
is to thermalize the current state before computing its contribution to the
averages. This can be done by performing a few canonical sweeps under the
temperature $T(E)$ corresponding to the current energy level. A rough
estimate for this value can be measured at the current state itself, through

$${1 \over T(E)} =
{1 \over \Delta{E}} \ln {N_{{\rm up},S}^{(\Delta{E})}
\over N_{{\rm dn},S}^{(\Delta{E})}} \,\,\,\, ,
\eqno(11)$$

\noindent or, again using the already quoted dirty trick,

$${1 \over T(E)} =
\ln {\sum_{\Delta{E}} (N_{{\rm up},S}^{(\Delta{E})})^{1/\Delta{E}}
\over \sum_{\Delta{E}} (N_{{\rm dn},S}^{(\Delta{E})})^{1/\Delta{E}}}
\,\,\,\, .
\eqno(12)$$

\noindent Some few such extra thermalization sweeps were adopted in
references [6,7], in between two RW sweeps. We observed that one RW sweep
followed by one canonical extra sweep is enough to eliminate the correlations
within our numerical accuracy.

Introducing extra canonical simulational steps into a method whose main
purpose is to eliminate some problems appearing in canonical simulations,
however, is not fair. Nevertheless, our main purpose in references [6,7] was
to test whether our fundamental equation (8) is valid or not, because we have
not yet proven that it is exact at this time. Now it is proven in general. On
the other hand, as already quoted, which particular simulational approach
would be adopted in order to calculate the microcanonical averages appearing
in (8) is a matter of user's choice. I have obtained very accurate results
[9] by adopting a microcanonical simulator [10] instead of the RW dynamics,
nothing to do with neither canonical simulations nor RW dynamics.

Finally, I show that RW dynamics introduced in [8] is not so bad as claimed
in [11], provided the proper corrections were made into its wrong (in
thermodynamic grounds) dynamic rule. The rule in [8] is: 1) compute the
numbers $N_{{\rm up},S}$ and $N_{{\rm dn},S}$ of possible movements one could
perform at the current state $S$, respectively increasing or decreasing its
energy; 2) choose one random movement and perform it according to a
probability $p_{\rm dn}$ proportional to $N_{{\rm up},S}$ in case this movement
decreases the energy, or $p_{\rm up}$ proportional to $N_{{\rm dn},S}$ in case
the energy would increase; 3) non-varying-energy movements can be performed
at will. According to this rule [8], the probability of performing some
tossed movement does not depend on the value of the energy jump $\Delta{E}$.
In order to measure thermal averages, this rule is certainly bad, once
different energies are treated within the same probability. This fundamental
concept was ignored in [11], in spite of being already noted many times
[12--15] within microcanonical simulations. In [11], of course, wrong
averages were found. However, the purpose of reference [8] is not to compute
thermal averages, but only to search for cost-minimum states in complex
landscapes. For this purpose, thermodynamic concerns do not matter (although
can help sometimes), and this dynamics works very well.

The purpose of reference [6] was to compute thermal averages. Namely, the
microcanonical averages of $N_{{\rm up},S}^{(\Delta{E})}$ and $N_{{\rm
dn},S}^{(\Delta{E})}$. Note the difference between these quantities and that
used in [8] (last paragraph): now, as already mentioned after equation (8),
both quantities correspond to a single value of $\Delta{E}$. Only movements
corresponding to the same energy jump $\pm\Delta{E}$ appear into the
microcanonical averages $<N_{\rm up}(E)>$ and $<N_{\rm dn}(E+\Delta{E})>$. In
order to calculate these averages, the rule in [6] is: 1) choose one random
movement, and measure its energy jump $\pm\Delta{E}$; 2) compute the numbers
$N_{{\rm up},S}^{(\Delta{E})}$ and $N_{{\rm dn},S}^{(\Delta{E})}$ of possible
movements one could perform at the current state $S$, respectively increasing
or decreasing its energy by the same amount $\Delta{E}$ of the tossed
movement; 3) perform it according to a probability $p_{\rm dn}$ proportional
to $N_{{\rm up},S}^{(\Delta{E})}$ in case this movement decreases the energy,
or $p_{\rm up}$ proportional to $N_{{\rm dn},S}^{(\Delta{E})}$ in case the
energy would increase; 4) non-varying energy movements can be performed at
will. Considering the particular case of the square lattice Ising ferromagnet
within the further particularity of the single-spin-flip protocol, one has
two different possible values for $\Delta{E}$. One corresponds to spins
surrounded by four parallel neighbours ($i = 2$, according to the notation
adopted in [11]), which must be flipped according to some probability, say
$y$. The other corresponds to spins surrounded by three parallel neighbours,
with a single one pointing in the opposite sense ($i = 1$), which must be
flipped according to probability $x$. The correct RW dynamics [6] corresponds
to take $y = x^2$, and not $y = x$ as in [11].

The table shows the wrong RW results (WRW) obtained in [11] for $N_i$, with
$i = -2$, $-1$, $1$ and $2$ (we use here the same notation), for the same $80
\times 80$ square lattice Ising ferromagnet, compared with the correct
canonical averages (CS, also copied from [11]). Now, I included the correct
results I got by using the correct RW dynamics (RW) introduced in [6]. I
computed only $32 \times 200$ states per energy level around the desired
region. This corrected RW rule [6], different from the one used in [11], is
good for computing thermal averages, as already tested in [6,7].
Nevertheless, I need to emphasize once again that this corrected RW rule is
not supposed to be confused with the Broad Histogram Method, i.e. equation
(8), for which any other microcanonical simulator can be applied. The
microcanonical averages appearing in equation (8) could even be calculated
from another protocol of allowed movements, completely distinct from that
used in order to count the numbers $N_{{\rm up},S}^{(\Delta{E})}$ and
$N_{{\rm dn},S}^{(\Delta{E})}$ at the current state.

Even within the corrected RW dynamics [6], some precautions in dealing with
the correlations may be important. As already quoted, the extra correlations
within RW dynamics, as compared to canonical, fixed temperature simulations,
are a consequence of the large energy jumps. Thus, an obvious and simple
way
to avoid these correlations is just to forbid these jumps. This can be done
by dividing the energy axis in adjacent small windows, performing RW dynamics
inside each window, sequentially. I have done this for the square lattice
Ising ferromagnet, and the results for the average energy and specific heat
are displayed in the figure. This is a much more crucial test than that
presented in [11] for a single energy value (see table), once it depends on
the whole function $g(E)$, along the whole energy axis. As one can see, the
quality of the results is the same as obtained first in [7] where canonical
steps were introduced in between RW steps, and [9], where another completely
different simulational dynamics [10] was adopted. All the three cases,
however, share the same status: they were obtained by using the exact
equation (8), more specifically its approximation (10).

Concluding, the Broad Histogram Method [6], equation (8), was proven to be
exact for any statistical model, or any energy spectrum, under completely
general conditions. It serves for determining the energy degeneracies $g(E)$
from the microcanonical averages $<N_{\rm up}(E)>$ and $<N_{\rm dn}(E)>$
measured at constant energies $E$.  The number $N_{\rm up}$ counts the
possible modifications, or movements one can perform at the current state
increasing its energy by an amount $\Delta{E}$. Analogously, $N_{\rm dn}$
counts the number of possible movements decreasing its energy by the same
amount.

I am indebted to my collaborators Thadeu and Hans, and also to the people
from the Metallurgy Engineering School in Volta Redonda, another campus of
my university where I give classes during this term, and where the proof of
equation (8) came to my mind.

\vskip 30pt
\centerline{\bf References}\parG
\item{[1]} N. Metropolis, A.W. Rosenbluth, M.N. Rosenbluth, A.H. Teller and
E. Teller, {\it J. Chem. Phys.} {\bf 21}, 1087 (1953).\par
\item{[2]} Z.W. Salzburg, J.D. Jacobson, W. Fickett and W.W. Wood, {\it J.
Chem. Phys.} {\bf 30} 65 (1959).\par
\item{[3]} A.M. Ferrenberg and R.W. Swendsen, {\it Phys. Rev. Lett.} {\bf
61}, 2635 (1988).\par
\item{[4]} E.P. M\"unger and M.A. Novotny, {\it Phys. Rev.} {\bf B43}, 5773
(1991).\par
\item{[5]} J. Lee, {\it Phys. Rev. Lett.} {\bf 71}, 211 (1993); B. Hesselbo
and R.B. Stinchcombe, {\it Phys. Rev. Lett.} {\bf 74}, 2151 (1995).\par
\item{[6]} P.M.C. de Oliveira, T.J.P. Penna and H.J. Herrmann, {\it Braz. J.
of Physics} {\bf 26}, 677 (1996); also in COND-MAT 9610041.\par
\item{[7]} P.M.C. de Oliveira, T.J.P. Penna and H.J. Herrmann, {\it Eur.
Phys. J.} {\bf B1}, 205 (1998).\par
\item{[8]} B. Berg, {\it Nature}, {\bf 361}, 708 (1993).\par
\item{[9]} P.M.C. de Oliveira, {\it Int. J. Mod. Phys.} {\bf C9}, 497
(1998).\par
\item{[10]} M. Creutz, {\it Phys. Rev. Lett.} {\bf 50}, 1411 (1983).\par
\item{[11]} B. Berg and U.H.E. Hansmann, {\it Eur. Phys. J.} {\bf B} (1998),
to appear.\par
\item{[12]} W.M. Lang and D. Stauffer, {\it J. Phys.} {\bf A20}, 5413
(1987).\par
\item{[13]} C. Moukarzel, {\it J. Phys.} {\bf A22}, 4487 (1989).\par
\item{[14]} P.M.C. de Oliveira, T.J.P. Penna, S.M. Moss de Oliveira and R.M.
Zorzenon, {\it J. Phys.} {\bf A24}, 219 (1991).\par
\item{[15]} P.M.C. de Oliveira, {\sl Computing Boolean Statistical Models},
World Scientific, Singapore (1991).\par
\item{[16]} P.D. Beale {\it Phys. Rev. Lett.} {\bf 76}, 78 (1996).\par

\vskip 30pt
\centerline{\bf Figure Caption}\parG
\item{Figure 1} Average energy (dots) and specific heat (crosses) for the $32
\times 32$ square lattice Ising ferromagnet. The continuous lines show the
exactly known curves [16], the specific heat peak also blown up in the inset.
This is a very rough estimate, with only $32 \times 1920$ Monte Carlo sweeps
along the whole energy axis, running in less than 6 minutes on a
workstation.\par

\vskip 20pt
\centerline{\bf Table Caption}\parG
\settabs 5\columns
\+\hskip 45pt $i$&\hskip 20pt $-2$&\hskip 20pt $-1$&\hskip 20pt $1$&\hskip
20pt $2$\cr
\+\hskip 40pt CS&$0.018853 (03)$&$0.072752 (04)$&$0.331070 (11)$&$0.389694
(09)$\cr
\+\hskip 35pt WRW&$0.034282 (26)$&$0.057936 (19)$&$0.350240 (43)$&$0.388130
(55)$\cr
\+\hskip 40pt RW&$0.018835$&$0.072766$&$0.331080$&$0.389678$\cr
\parG
\item{Table 1} To be compared with Table 1 of reference [11], where the wrong
results (WRW) obtained from the wrong RW dynamics are now corrected by using
the correct RW dynamics (RW) introduced in [6]. The error bars are
supposed to be similar to that of WRW. Correct canonical simulations
(CS) [11] are also included for comparison.\par
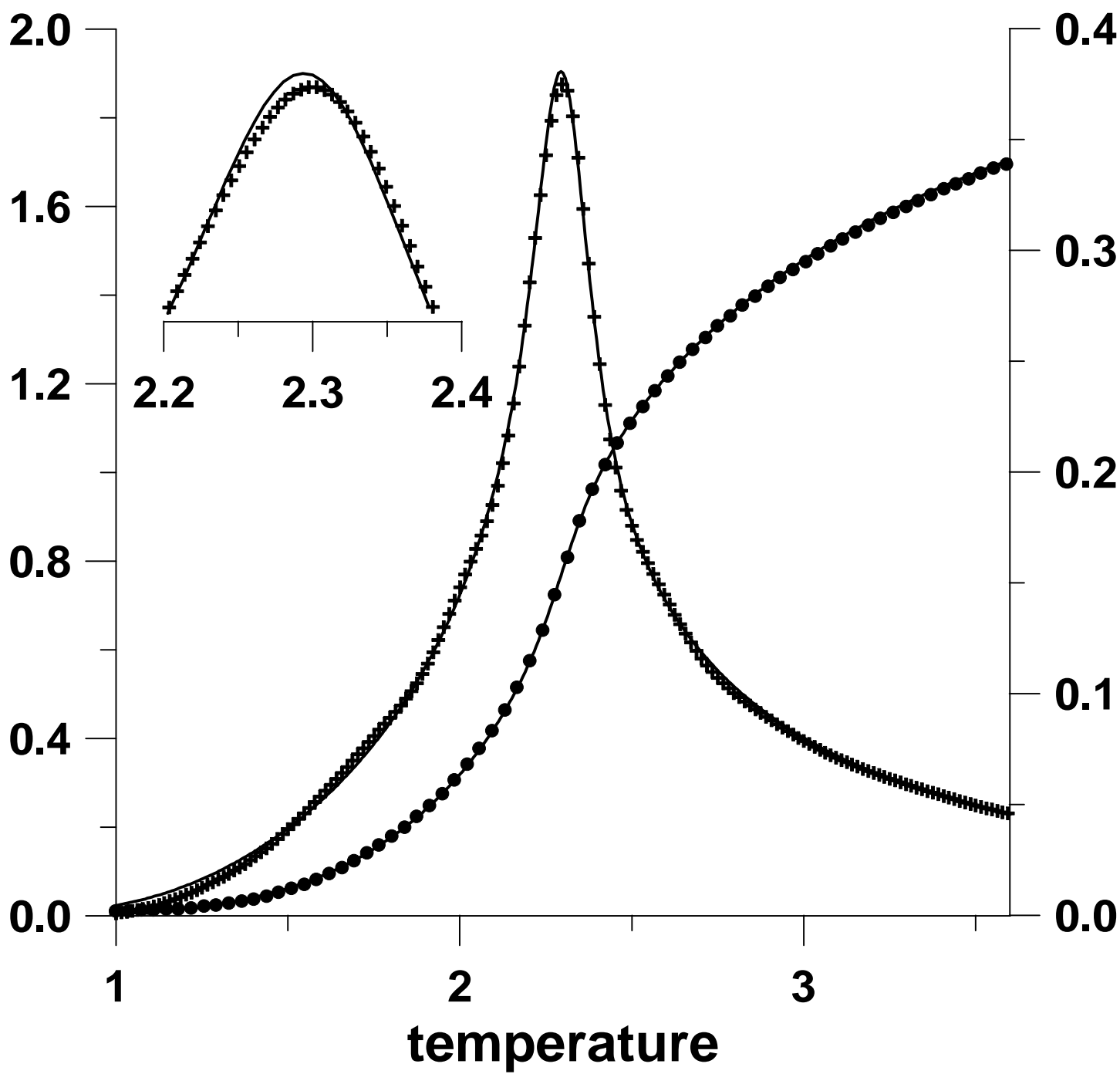
\bye